\documentclass[aps,twocolumn
,superscriptaddress,noshowpacs,nofootinbib
,tightenlines]{revtex4-1}
\usepackage{mathrsfs}
\usepackage{latexsym,bm}
\usepackage{graphicx}
\usepackage{indentfirst}
\usepackage{slashed}
\usepackage{amssymb}
\usepackage{amsmath}
\usepackage{bbm}
\usepackage{color}
\usepackage[dvipsnames]{xcolor}
\usepackage{epsfig}
\usepackage[titletoc]{appendix}
\usepackage{multirow}%
\usepackage{rotating}
\usepackage{epstopdf}
\usepackage{extarrows}
\usepackage{tabularx}
\usepackage{soul} 
\usepackage{xcolor}
\usepackage{lipsum}
\usepackage[colorlinks=true,allcolors=BlueViolet,hyperfootnotes=false]{hyperref}

\usepackage[utf8]{inputenc}

\newcommand{\be}{\begin{equation}} \newcommand{\ee}{\end{equation}}
\newcommand{\ba}{\begin{array}{c}} \newcommand{\ea}{\end{array}}
\newcommand{\bea}{\begin{eqnarray}} \newcommand{\eea}{\end{eqnarray}}

\begin{document}
\title{\Large Further tests of lepton
  flavour universality from the charged lepton energy distribution in $b\to c$ semileptonic decays: The case of $\Lambda_b\to \Lambda_c \ell \bar\nu_\ell$}

\author{Neus Penalva}
\affiliation{Instituto de F\'{\i}sica Corpuscular (centro mixto CSIC-UV), Institutos de Investigaci\'on de Paterna,
Apartado 22085, 46071, Valencia, Spain}

\author{Eliecer Hern\'andez}
\affiliation{Departamento de F\'\i sica Fundamental 
  e IUFFyM,\\ Universidad de Salamanca, E-37008 Salamanca, Spain}

\author{Juan~Nieves}
\affiliation{Instituto de F\'{\i}sica Corpuscular (centro mixto CSIC-UV), Institutos de Investigaci\'on de Paterna,
Apartado 22085, 46071, Valencia, Spain}

\date{\today}
\begin{abstract}
In a general framework, 
valid for any $H\to H'\ell^-\bar\nu_\ell$ 
semileptonic decay, we analyze the   
$d^2\Gamma/(d\omega d\cos\theta_\ell)$ and $d^2\Gamma/
(d\omega dE_\ell)$ distributions, with  $\omega$  being the product of the hadron four-velocities,  $\theta_\ell$  the angle made 
by the three-momenta of 
the charged lepton and the final hadron in the $W^-$ center of mass frame and 
$E_\ell$  the charged lepton energy in the decaying hadron
rest frame. Within the Standard Model (SM), $ d^2\Gamma/(d\omega dE_\ell)\propto 
\left(c_0(\omega)+c_1(\omega)E_\ell/M+c_2(\omega)E^2_\ell/M^2\right)$, with $M$
the initial hadron mass. We find that   $c_2(\omega)$ is 
independent of the lepton flavor and thus it is an ideal candidate to look for lepton
flavor universality (LFU) violations.  We also
find a correlation between the $a_2(\omega)$ structure function, that governs the
$(\cos\theta_\ell)^2$ dependence of $d^2\Gamma/(d\omega d\cos\theta_\ell)$, and
$c_2(\omega)$.  Apart from trivial kinematical and mass factors, the ratio of 
$a_2(\omega)/c_2(\omega)$ is a universal function that 
can be measured in any semileptonic decay, involving not only $b\to c$ 
transitions. These two  SM predictions  can be used as new tests in the present search for  
signatures of LFU violations. We also generalize the formalism to account for some new physics (NP) terms, 
and show that neither $c_2$ nor $a_2$ are 
 modified by left and right  scalar NP terms, being  however sensitive to  left and right
 vector  corrections. We also find that the  $a_2/c_2$ ratio 
 is not modified by these latter NP contributions. Finally, and in order to illustrate our findings, 
we apply our general framework to the $\Lambda_b\to \Lambda_c \ell \bar\nu_\ell$ decay.
We show that a  measurement of $c_2$ (or $a_2$) for $\tau$ decay would not  
only be a direct measurement of the possible existence of NP, but it would also allow to
distinguish from NP fits to $b\to c\tau\bar\nu_\tau$ anomalies in the meson sector, that otherwise give the same  total and differential $d\Gamma/d\omega$ widths.
We show that the same occurs for the other two terms, $c_0$ and $c_1$, that appear in 
$d^2\Gamma/(d\omega d E_\ell)$, and for the $\cos\theta_\ell$ linear term of the angular distribution.

\end{abstract}

\maketitle
\section{Introduction}
%
The discrepancies, between  available data and the Standard Model (SM) predictions seen 
 in semileptonic $B-$meson decays, point at the possible existence of new physics (NP), 
 affecting to  the third quark and lepton generations, 
 responsible for  lepton flavor universality (LFU) violations (for a recent review see 
 Ref.~\cite{Bifani:2018zmi}).  Present average
results for the  ${\cal R}_{D^{(*)}}=\frac{\Gamma(B\to D^{(*)}\tau\bar\nu_\tau)}{\Gamma(B\to
D^{(*)}\ell\bar\nu_\ell)}$ ratios ($\ell=e,\mu$), show a tension with the SM predictions at 
the  4.4 standard deviations ($\sigma$) level [Heavy Flavour Averaging Group (HFLAV)~\cite{Amhis:2016xyh}, using BaBar~\cite{Lees:2012xj,Lees:2013uzd}, Belle~\cite{Huschle:2015rga,Sato:2016svk,Hirose:2016wfn} and LHCb~\cite{Aaij:2015yra,Aaij:2017uff} data and SM predictions~\cite{Aoki:2016frl,Aaij:2015yra, Bigi:2016mdz, Jaiswal:2017rve} ]. New preliminary measurements by the Belle
Collaboration~\cite{Abdesselam:2019dgh} reduce however this tension with the SM predictions to 1.2$\sigma$. 
A general model-independent analysis of different $b\to c \ell \nu_\ell$ charged current 
(CC) transition operators has been addressed in Ref.~\cite{Murgui:2019czp} within an effective field theory
approach. The main conclusion of this study is that the anomaly is still present and can be
solved by NP, in agreement with previous works (see f.i. the pioneering work of Ref.~\cite{Fajfer:2012vx}).

This anomaly can be corroborated  in
$\Lambda_b\to\Lambda_c\ell\bar\nu_\ell$ decays, which are also governed by the $b\to c$ transition. The $\omega-$shape of the  
differential width for muons  has been recently measured by the LHCb Collaboration~\cite{Aaij:2017svr}, and there exist
prospects~\cite{Cerri:2018ypt} that the level of precision in the ${\cal
R}_{\Lambda_c}=\frac{\Gamma(\Lambda_b\to\Lambda_c\tau\bar\nu_\tau)}{
\Gamma(\Lambda_b\to\Lambda_c\mu\bar\nu_\mu)}$ ratio might
reach that obtained for ${\cal R}_{D^{(*)}}$. The form factors relevant for this
transition are strongly constrained by heavy quark
spin symmetry (HQSS), since no subleading Isgur-Wise (IW) function occurs at order ${\cal O}(\Lambda_{QCD}/m_{b,c})$ and only two 
subleading functions enter at the next order
 \cite{Neubert:1993mb,Bernlochner:2018kxh}.  Precise results for the form factors 
  were obtained in  Ref.~\cite{Detmold:2015aaa} using Lattice
QCD (LQCD) with 2+1 flavors of dynamical domain-wall fermions. Leading and subleading HQSS IW functions are simultaneously fitted to LQCD 
results and LHCb data,  and are used to accurately predict  the ${\cal
R}_{\Lambda_c}$ ratio in the SM~\cite{Bernlochner:2018kxh}.
Therefore, this reaction is, from the theoretical point of view, 
as appropriate as the $B\to D^{(*)}$ processes for the study of $b\to c$ LFU 
violations. A sum rule relating 
${\cal R}_{\Lambda_c}$ to ${\cal R}_{D^{(*)}}$, independent of
any NP scenario up to small corrections, was
found in Refs.~\cite{PhysRevD.99.075006,PhysRevD.100.035035}. There it is shown
that ${\cal R}_{\Lambda_c}$ does not provide additional information on the
Lorentz structure of NP but provides an important consistency
check of the ${\cal R}_{D^{(*)}}$ measurements. The full four-differential 
 angular distribution of the
 $\Lambda_b\to\Lambda_c(\to \Lambda^0\pi^+)\ell^-\bar\nu_\ell$ decay 
 has been recently studied in Ref.~\cite{Boer:2019zmp} with the finding that the
 full set of angular observables analyzed is sensitive to more combinations of 
 NP couplings than the ${\cal R}_{D^{(*)}}$ ratios. In this latter reference some discrepancies with the results 
 of the previous study of Ref.~\cite{PhysRevD.91.115003} are pointed out. NP corrections to 
${\cal R}_{\Lambda_c}$ have also been examined in other works~\cite{Li:2016pdv, Datta:2017aue, Bernlochner:2018bfn,Ray:2018hrx, 
 DiSalvo:2018ngq, Murgui:2019czp,Ferrillo:2019owd}. Some of them pay also attention to the double differential rate, 
 $d^2\Gamma/(d\omega d\cos\theta_\ell)$, in addition to ${\cal R}_{\Lambda_c}$ or the $\omega$ spectrum ($\omega$  being the 
 product of the hadron four-velocities and  $\theta_\ell$  the angle made  by the three-momenta of the charged lepton and the
  final hadron in the  center of mass (CM) of the two final leptons).  Thus, forward-backward asymmetry has been  calculated
   for this baryon decay~\cite{Li:2016pdv, Datta:2017aue, Bernlochner:2018bfn, Ray:2018hrx}, while the full CM charged 
   lepton angular dependence has also been  analyzed for $B-$meson reactions, see f.i.~\cite{Fajfer:2012vx,Murgui:2019czp}. 
   However, to our knowledge, the charged lepton energy ($E_\ell$) distribution in the decaying hadron
rest frame  has never been considered neither for   $\Lambda_b\to\Lambda_c$ nor $B\to D^{(*)}$ semileptonic  decays.

Finally, we should also  mention that the $\Lambda_b\to\Lambda_c\ell\bar\nu_\ell$ decay  provides an alternative
 method to determine the  Cabibbo-Kobayashi-Maskawa matrix element $|V_{cb}|$ and to study the unitarity  triangle within the SM.

In this work we introduce a general framework to study any baryon/meson 
semileptonic decay for unpolarized hadrons, though we refer explicitly here to those induced by the $b\to c$ transition.
 Within this scheme we  find
general expressions for the $d^2\Gamma/(d\omega d\cos\theta_\ell)$ and $d^2\Gamma/
(d\omega dE_\ell)$ differential decay widths (see below), each of them expressible in terms of 
three different structure functions (SFs). Proceeding in this way, we have uncovered two
new observables that can be measured  and  used as model independent tests
for LFU violation analyses. This is discussed in next section, and it 
constitutes the most relevant result of this work. Indeed, we identify two contributions,  one in the 
$E_\ell$ spectrum  and a second one in the $\cos\theta_\ell$ distribution, 
which are independent of the lepton flavor in the SM. They provide  novel, 
model-independent and clean  tests of LFU. Moreover, we show that the ratio of both 
of them  within the SM should be a universal function, which could be measured in all
type of hadron (baryon or meson) semileptonic decays governed by CC $c\to s$, $c\to d$, $s\to u$, $b\to u$ transitions. This charged lepton energy-angle  correlation should be experimentally accessible and, if violated, it would be a
 clear indication of NP, also eliminating  some  possible Lorentz structures for the new terms as we will discuss.

 We also generalize the formalism to account for left and right  scalar and vector NP contributions using the scheme of Ref.~\cite{Murgui:2019czp}.
 
 To illustrate our findings,  we will apply the general framework   
 to the analysis of the $\Lambda_b\to\Lambda_c$ decay. Using the state of the art LQCD form factors of 
Ref.~\cite{Detmold:2015aaa}, we evaluate the six SM SFs and the
$d\Gamma/d\omega$ differential rate. For the case of a final $\tau$ lepton we give explicitly the contributions coming from  positive and negative helicities measured, both in the CM of the $W^-$  boson and in the $\Lambda_b$ rest frame (LAB).  As mentioned above, the relevance  of the LAB lepton energy spectrum 
for LFU violation has never been studied for $\Lambda_b$ or for $B$ decays. Indeed, we also discuss how some features of that spectrum can be used 
to distinguish between different NP scenarios  that otherwise lead to the same  total and differential 
semileptonic decay widths.

\section{Decay Width}
We consider the  semileptonic decay of a bottomed hadron
($H_b$) into a charmed one ($H_c$) and $\ell \bar\nu_\ell$, driven by the CC $b\to c$
 transition. In the SM, the differential decay width for massless neutrinos reads~\cite{Tanabashi:2018oca},
\begin{equation}
  \frac{d^2\Gamma}{d\omega ds_{13}} = \frac{G^2_F|V_{cb}|^2
    M'^2}{(2\pi)^3 M} L^{\mu\nu}(k,k')W_{\mu\nu}(p,q)
  \end{equation}
with $G_F=1.166\times 10^{-5}$~GeV$^{-2}$  the Fermi coupling
constant, $M$ ($M'$) the mass of the initial (final) hadron and $W$
and $L$, the hadron and lepton tensors. The latter one, after summing
over all lepton polarizations is given by ($\epsilon_{0123}=+1$)
\begin{equation}
  L_{\mu\nu}(k,k')= k'_{\mu} k_{\nu}+k_{\mu} k'_{\nu} - g_{\mu\nu} k\cdot k'+ i\epsilon_{\mu\nu\alpha\beta}k'^{\alpha} k^{\beta}\, ,
 \end{equation}  
with $k'$ ($k$) the outgoing charged lepton (neutrino)
four-momentum. In addition, the product of the two hadron four velocities $\omega$
and $q^2=(k+k')^2$ are related via $q^2 = M^2+M'^2-2MM'\omega$ and
$s_{13}=(p-k)^2$, with $p$ the four momentum of the decaying $H_b$
particle. Finally, the dimensionless hadron tensor is constructed from the
non-leptonic CC vertex $j^\mu_{cc}=\bar c(0) \gamma^\mu(1-\gamma_5)b(0)$  as
\begin{eqnarray}
  W^{\mu\nu}(p,q) = \overline{\sum} \langle H_c;  p' |
  j_{cc}^\mu | H_b; p\rangle \langle H_c; p' |
  j_{cc}^\nu | H_b; p \rangle^* \label{eq:defWmunu}
  \end{eqnarray}
with $p'= p-q$. The sum is done over initial (averaged) and final hadron
spins, and the states are normalized as $\langle \vec{p}, r| \vec{p}\,',
s\rangle= (2\pi)^3(E/M)\delta^3(\vec{p}-\vec{p}\,')\delta_{rs}$, with $r,s$  spin
indexes. Lorentz covariance leads to the general decomposition 
\begin{eqnarray}
W^{\mu\nu}(p,q)&=&-g^{\mu\nu}W_1+\frac{p^{\mu}p^{\nu}}{M^2}W_2+i\epsilon^{\mu\nu\alpha\beta}p_{\alpha}q_{\beta}\frac{W_3}{2M^2}\nonumber
\\
&+&\frac{q^{\mu}q^{\nu}}{M^2}W_4+\dfrac{p^{\mu}q^{\nu}+p^{\nu}q^{\mu}}{2M^2}W_5 \label{eq:wmunu-dec}
\end{eqnarray}
actually valid for any $H\to H'$
 CC transition with unpolarized hadrons\footnote{We have not included an
antisymmetric term proportional to $(p^{\mu}q^{\nu}-p^{\nu}q^{\mu})$,
since it would lead to time-reversal odd
correlations~\cite{Hernandez:2007qq, Sobczyk:2018ghy}.}.
The $W_i$ SFs are scalar functions of $q^2$ or
equivalently of $\omega$.
The double differential decay width can be rewritten
introducing the angle ($\theta_\ell$) made by the charged lepton ($\ell$) 
and the final hadron
($H_c$) in the $W^-$boson CM frame as
\begin{eqnarray}
  \frac{d^2\Gamma}{d\omega d\cos\theta_\ell}& =& \frac{G^2_F|V_{cb}|^2
    M'^3M^2}{16\pi^3}
  \sqrt{\omega^2-1}\left(1-\frac{m_\ell^2}{q^2}\right)^2
  A(\theta_\ell) \nonumber \\
  A(\theta_\ell) &=& a_0(\omega)  + a_1(\omega) \cos\theta_\ell +
  a_2(\omega) \cos^2\theta_\ell \label{eq:distr-ang} \\
  a_0 &=&2\frac{q^2}{M^2}W_1+ \frac{M'^2}{M^2}(\omega^2-1)W_2 \nonumber \\
      &+&\frac{m^2_\ell}{q^2}\left(
 \frac{M_\omega^2}{M^2}W_2+\frac{q^4}{M^4}W_4+\frac{q^2M_\omega}{M^3}W_5\right)\nonumber \\
  a_1 &=& -\frac{q^2M'}{M^3}\sqrt{\omega^2-1}\, W_3 \nonumber \\
      &+&\sqrt{\omega^2-1}\,\frac{m^2_\ell}{q^2}\frac{M'}{M}\left(2\frac{M_\omega}{M} W_2+\frac{q^2}{M^2}W_5\right)\nonumber \\
  a_2&=& -\frac{M'^2}{M^2} (\omega^2-1) \left( 1-
  \frac{m^2_\ell}{q^2}\right)W_2  \label{eq:as}
\end{eqnarray}
with $M_\omega=(M-M'\omega)$ and
$m_{\ell}$ the mass of the charged lepton. The variable $\omega$
varies from 1 to $\omega_{\rm max}= (M^2+M'^2-m^2_{\ell})/(2MM')$ and
$\cos\theta_\ell$ between $-1$ and 1. The terms proportional to $m_\ell^2$ in each of the 
coefficients $a_{i=0,1,2}$ account for the contributions from positive
helicity of the outgoing $\ell$. This  follows from the
expression of the  lepton tensor for a charged lepton with well defined helicity ($h=\pm 1$),
\begin{eqnarray}
 L_{\mu\nu}(h) &=&
 \frac{L_{\mu\nu}}{2}-  \frac{h}{2} s^\alpha \nonumber\\
 &\times& \left( k_\mu g_{\nu\alpha}+ k_\nu g_{\mu\alpha}-k_\alpha g_{\mu\nu}+i \epsilon_{\mu\nu\alpha\beta}k^\beta\right)\, ,
 \label{eq:lept_pol}
\end{eqnarray}
where $s^\alpha= (|\vec{k}'|, k^{\prime 0}\hat k')$ and 
$\hat k'=\vec{k}'/|\vec{k}'|$.

We are also interested in the double differential decay width with respect to 
$\omega$ and the energy 
($E_\ell$)  of the charged lepton in the LAB frame
\begin{eqnarray}
  \frac{d^2\Gamma}{d\omega dE_\ell}& =& \frac{G^2_F|V_{cb}|^2
    M'^2M^2}{8\pi^3}\left[ c_0(\omega)  + c_1(\omega) \frac{E_\ell}{M} +
  c_2(\omega)  \frac{E^2_\ell}{M^2}\right] \nonumber \\
  c_0 &=&\frac{q^2}{M^2}\left(2W_1-W_2 - \frac{M_\omega}{M}W_3\right) \nonumber \\
      &+&\frac{m^2_\ell}{M^2}\left(-2W_1+W_2+
  \frac{M_\omega}{M}(2W_4+2W_5-W_3)\right. \nonumber \\
      &-&\left. \frac{M^2-M'^2+m^2_\ell}{M^2}W_4\right)\nonumber \\
  c_1 &=& 2 \left(2 \frac{M_\omega}{M}W_2 +\frac{q^2}{M^2}W_3-\frac{m_\ell^2}{M^2}W_5\right)\nonumber \\
  c_2&=& -4 W_2 \label{eq:distr-lab}
\end{eqnarray}
with $E_\ell\in[E_\ell^{-},E_\ell^{+}]$ and
\begin{equation}
  E_\ell^{\pm}= \frac{M_\omega(q^2+m^2_\ell)  \pm M' \sqrt{\omega^2-1}(q^2-{m^2_\ell})}{2q^2}
\end{equation}  
The relevance of this distribution is in the fact  that within the SM, and up to small electroweak
  corrections, the $c_2(\omega)$ SF does not depend on the lepton mass.  
Therefore,  this function determined in $\ell=e, \mu$ decays should be the same as
that seen in $\tau$ decays. This is a clear test for LFU in
all type of semileptonic $b\to c$ decays that to our knowledge has not been
considered so far.  A similar comment holds for $a_2(\omega)$
entering in the CM angular distribution of Eq.~\eqref{eq:distr-ang}, after accounting for the
trivial $(1-m_\ell^2/q^2)$ kinematical factor. Furthermore, the ratio
\begin{equation}
\frac{M^2}{M'^2}\frac{a_2(\omega)}{(1-m_\ell^2/q^2)c_2(\omega)} =  (\omega^2-1)/4
\label{eq:a2c2}
\end{equation}
is a universal function that  should be found in all
type of $q\to q'\ell\bar \nu_\ell$ transitions, since in that ratio the SF $W_2$ cancels out.
This is a  test of the predictions of the
SM and, in principle, this ratio can be measured in any semileptonic
decay:  $D\to \bar K$, $D\to \pi$, $D\to \bar K^*$, $D\to \rho$,
$\bar K\to \pi $, 
$\bar B_s\to  K$, $\bar B \to \pi$, $\Lambda_b\to \Lambda_c$, $\Lambda
 \to p$, $\Lambda_c\to \Lambda$, etc. Notice that a left-handed vector current
 that couples exclusively to the $\tau$ lepton, which is the so far  
 preferred NP
 explanation of the anomalies~\cite{Murgui:2019czp},  will have no effect
  on the ratio
 in Eq.~(\ref{eq:a2c2}). However, in that case both
 $a_2(\omega)/(1-m_\ell^2/q^2)$ and $c_2(\omega)$ will change for a final 
 $\tau$ by a factor
 $|1+C_{V_L}|^2$, where we follow here the notation in 
 Ref.~\cite{Murgui:2019czp}. A  right-handed vector current (the
 $C_{V_R}$ term in Eq.~(2.1) in Ref.~\cite{Murgui:2019czp}) will affect
  the $\omega$ dependence of both $a_2(\omega)/(1-m_\ell^2/q^2)$
  and $c_2(\omega)$ while the ratio in Eq.~(\ref{eq:a2c2}) will still go unaffected. 
  In this case one would expect $c_2$ and $a_2$ to change differently than the total decay width $\tau/\mu$ ratios.
  Further Lorentz dependencies, like the scalar and tensor ones, that include modifications
  in the lepton vertexes, would in principle modify all three quantities. We
  will further illustrate this point below in Subsec.~\ref{sec:c2a2}.
  In any case, 
  any violation of Eq.~(\ref{eq:a2c2}) will be a clear indication of NP beyond the SM, not driven by 
  left or right-handed vector current operators.

  Using Eq.~\eqref{eq:lept_pol}  in the LAB frame, we obtain  $d^2\Gamma(h)/(d\omega dE_\ell)$ 
for a charged lepton with a well defined helicity ($h=\pm 1)$
\begin{eqnarray}
  \frac{d^2\Gamma(h)}{d\omega dE_\ell}& =& \frac{G^2_F|V_{cb}|^2
    M'^2M^2}{8\pi^3} C_h(\omega,E_\ell) \\
   C_h &=& \left(\frac12- \frac{h}2
   \frac{E_\ell}{p_\ell}\right)\left[ c_0  + c_1 \frac{E_\ell}{M}+ c_2
     \frac{E^2_\ell}{M^2}\right]\nonumber \\
   &-&\frac{h}2 \frac{m^2_\ell}{ Mp_\ell}\left[\widehat{c}_0(\omega) +
     \widehat{c}_1(\omega) \frac{E_\ell}{M}\right] \\
   \widehat{c}_0(\omega) & = & -2 \frac{M_\omega}{M}\left(2
   W_1+W_2-\frac{m^2_\ell}{M^2}W_4\right) -
   \frac{q^2-m^2_\ell}{M^2}W_5\nonumber \\
   &+&\frac{M_\omega^2+(\omega^2-1)M'^2+m^2_\ell}{M^2} W_3 \\
   \widehat{c}_1(\omega) & = &2\left(2W_1+W_2-\frac{M_\omega}{M}W_3-\frac{q^2}{M^2}W_4\right)
\end{eqnarray}
with $p_\ell = \sqrt{E_{\ell}^2-m^2_\ell}$ the charged lepton three
momentum.  For a massless charged lepton
the $h=+1$ contribution vanishes, as expected from conservation of chirality. 

The individual contributions to 
$d^2\Gamma/(d\omega dE_\ell)$ from $\tau$ leptons with positive and negative
 helicity in the  LAB frame
can not be obtained from the depolarized $\ell=\mu, e$ 
and $\ell=\tau$ data alone. In contrast,  neglecting the electron or muon masses, the angular distribution of
Eq.~\eqref{eq:distr-ang} can be used, together with measurements of the
$\ell=\mu, e$ and $\ell=\tau$ $d^2\Gamma/(d\omega d\cos\theta_\ell)$
differential decay width, to separate the individual contributions of positive and negative
$\tau$ helicities in the CM frame. This is to say, with  great accuracy,
$(1-m_\tau^2/q^2)^{-2}\times d^2\Gamma/(d\omega d\cos\theta_\tau)$ for a $\tau$  
with negative helicity  can be determined from   the unpolarized $d^2\Gamma/(d\omega d\cos\theta_\ell)$  measured  
for muons or electrons.
\\

\section{Semileptonic $\Lambda_b^0\to  \Lambda_c^+\ell^-\bar\nu_\ell$ decay}
In this section we apply the above-described general formalism to the study of
the  semileptonic $\Lambda_b\to\Lambda_c $ decay. We present first SM results, and later we also discuss 
the effect of some NP contributions to the $a_2$ and $c_2$ coefficients.
\subsection{Form Factors}
The hadronic  matrix
element  can be parameterized in terms of three vector ($F_i$) and three axial
($G_i$)  form-factors, which are functions of $\omega$ and that are greatly
constrained by HQSS near zero recoil ($\omega=1$)~\cite{Neubert:1993mb,Bernlochner:2018kxh}
\begin{eqnarray}
\langle\Lambda_c;\vec{p}^{\,\prime};s|j^{\mu}_{cc}|\Lambda_b;\vec{p};r\rangle
&=& 
\bar{u}^{(s)}_{\Lambda_c}(\vec{p}^{\,\prime}\,)\Bigg\{\sum_{i=1}^3 {\cal
  O}^\mu_i{\cal F}_i\Bigg\} u^{(r)}_{\Lambda_b}(\vec{p}\,) \nonumber \\
{\cal  O}^\mu_1 = \gamma^{\mu}, \, {\cal  O}^\mu_2 &=&
\frac{p^{\mu}}{M_{\Lambda_b}}, \, {\cal  O}^\mu_3 =\frac{p^{\prime\mu}}{M_{\Lambda_c}}
\label{eq.FactoresForma}
\end{eqnarray}
with ${\cal F}_i=(F_i-\gamma_5G_i), \,i=1,2,3$ and  $u_{\Lambda_b,
  \Lambda_c}$ dimensionless Dirac spinors. In this case
\begin{eqnarray}
  W^{\mu\nu}(p,q)&=& \frac12 {\rm Tr}\Bigg[\frac{\slashed{p}'+M_{\Lambda_c}}{2M_{\Lambda_c}}\left(\sum_{i=1}^3 {\cal
  O}^\mu_i{\cal
      F}_i\right)\frac{\slashed{p}+M_{\Lambda_b}}{2M_{\Lambda_b}} \nonumber\\
    &\times&  \gamma^0\Bigg(\sum_{j=1}^3
    {\cal
  O}^\mu_j{\cal F}_j\Bigg)^\dagger\gamma^0\Bigg] \label{eq:wmunu-baryons}
\end{eqnarray}  
From this equation one  can obtain the $W_{i}$ SFs, and hence the $a_i,c_i$ coefficients, in terms of $F_i$ and $G_i$. 
The explicit expressions are given in  Appendix~\ref{app:ff}. These form factors (Eq.~\eqref{eq.FactoresForma}) are easily related to those used in 
the LQCD calculation of Ref.~\cite{Detmold:2015aaa} (see also  Appendix~\ref{app:ff}), which were given in terms of the Bourrely-Caprini-Lellouch
parametrization~\cite{Bourrely:2008za} (see Eq.~(79) of
\cite{Detmold:2015aaa}). A different determination of the form factors within
QCD sum rules in full theory is done in Ref.~\cite{PhysRevD.97.074007}. Taking into
 account the experimental
and theoretical uncertainties, the
LQCD form-factors  describe well the $\Lambda_b^0\to
  \Lambda_c^+\mu^-\bar\nu_\mu$ normalized spectrum
$(d\Gamma/dq^2)/\Gamma$   recently  measured by  the LHCb Collaboration~\cite{Aaij:2017svr}
   (see Fig.~5 in that reference).
From the integrated
distribution given in Ref.~\cite{Detmold:2015aaa} and using the $\Lambda_b^0$
lifetime ($1.471\pm 0.009$
ps) and  the $\Lambda_b^0\to \Lambda_c^+\mu^-\bar\nu_\mu$ branching
fraction [$(6.2\pm 1.4)\%$] quoted in ~\cite{Tanabashi:2018oca}, one obtains
$|V_{cb}|=0.044 \pm 0.005$ which is compatible with the values
reported by the HFLAV~\cite{Amhis:2016xyh}. 

For  numerical calculations  we use here the 11 parameters and statistical 
correlations given in Tables VIII and IX of Ref.~\cite{Detmold:2015aaa}. 

\subsection{SM results}
\begin{widetext}

\begin{figure}[tbh]
\includegraphics[height=4cm]{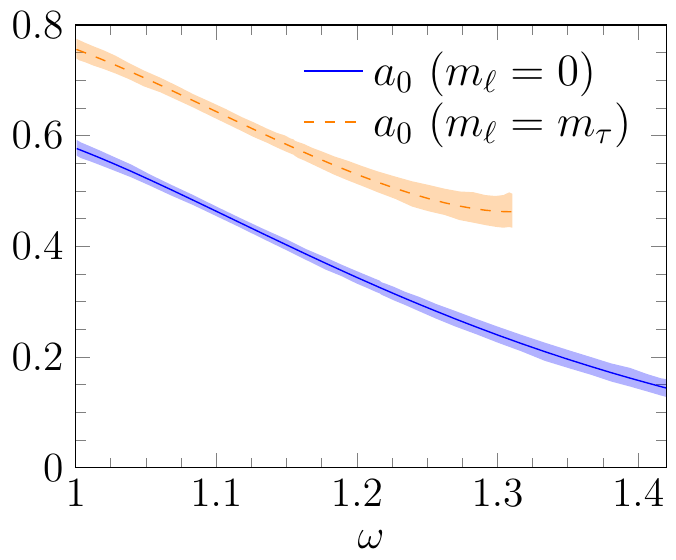}\ \includegraphics[height=4cm]{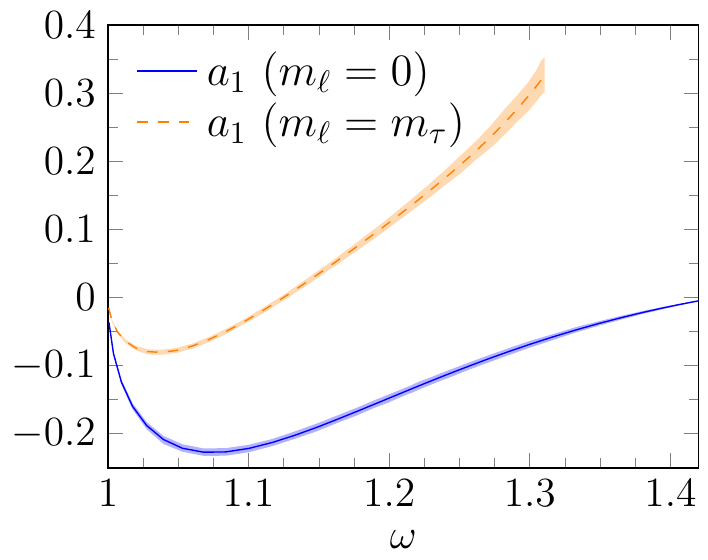}\ 
\includegraphics[height=4cm]{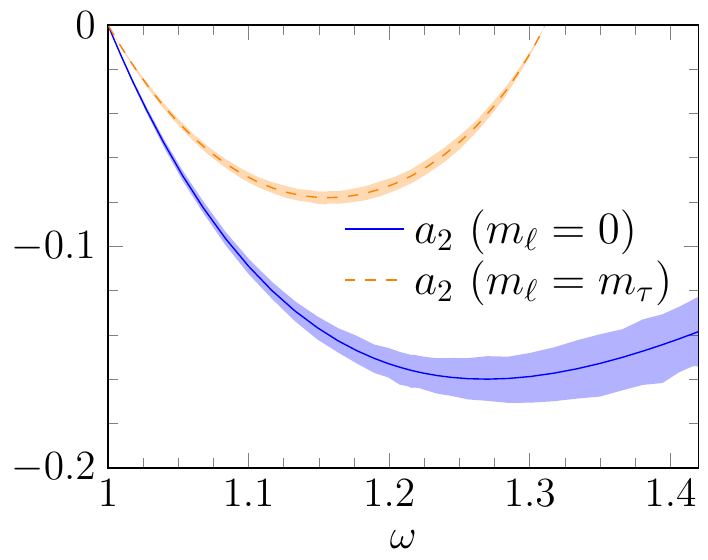}\caption{ Angular SFs $a_0,a_1$ and $a_2$ 
(Eq.~\eqref{eq:distr-ang}) for the $\Lambda_b^0\to
  \Lambda_c^+\ell^-\bar\nu_\ell$ decay obtained using the LQCD 
  form factors of Ref.~\cite{Detmold:2015aaa}. Bands account for 
  68\% CL intervals deduced from the correlation matrix given 
  in \cite{Detmold:2015aaa}.}  
  \label{fig:a012}
\end{figure}

\end{widetext}

\begin{widetext}

\begin{figure}[tbh]
\includegraphics[height=4cm]{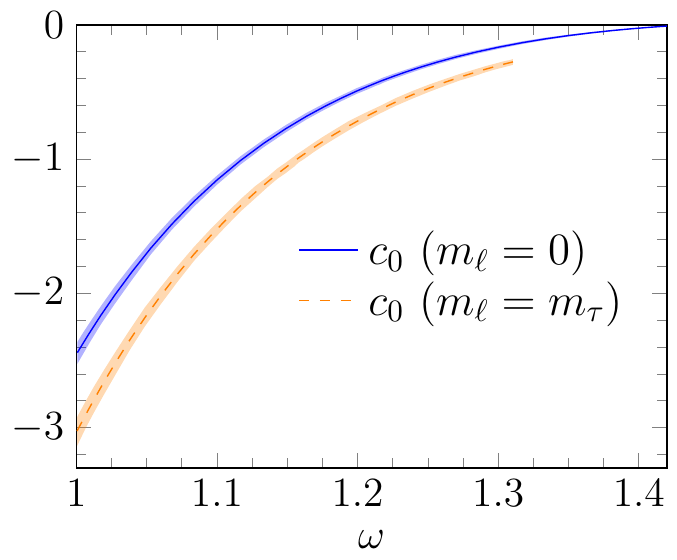}\ \includegraphics[height=3.825cm]{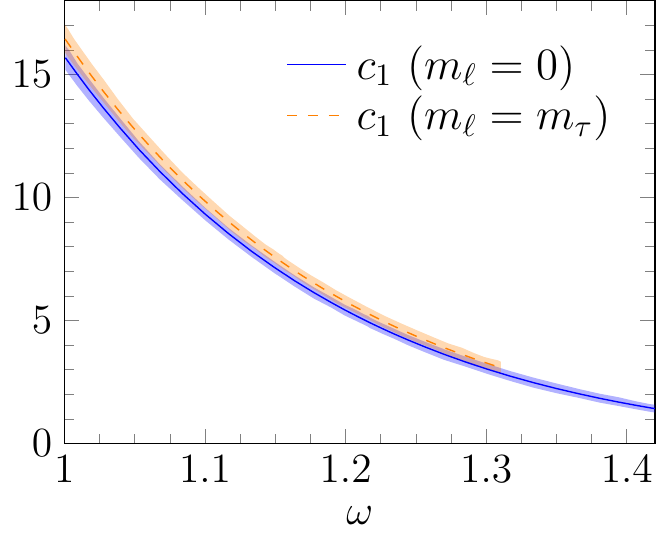}\ 
\includegraphics[height=4cm]{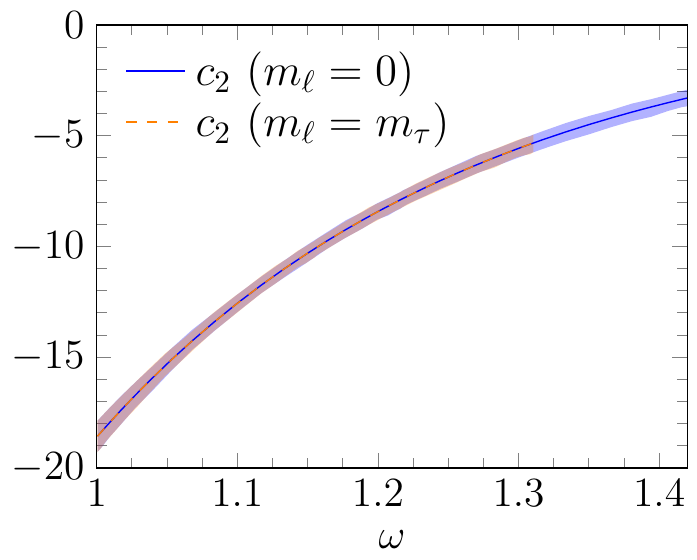}\caption{ The same as Fig.~\ref{fig:a012}, but for the $c_0,c_1$ and $c_2$ SFs defined in the LAB distribution of Eq.~\eqref{eq:distr-lab}.}  
  \label{fig:c012}
\end{figure}

\end{widetext}

The results 
obtained for the
$a_i,c_i$ SFs, both for $m_\ell=0$
(appropriate for $\ell=e,\mu$) and for $m_\ell=m_\tau$, are shown in
Figs.~\ref{fig:a012} and \ref{fig:c012}. We also display the 68\% confident level (CL) 
bands
that we Monte Carlo derive from the correlation matrix reported in \cite{Detmold:2015aaa}.
As mentioned above, within the SM, the $c_2(\omega)$ SF is the same 
for all charged leptons, providing  a new testing ground for  LFU violation
studies in $b\to c$ decays. We also observe that finite lepton mass corrections are quite small for $c_1$, while become more sizable for the rest of the SFs, which are given here for the very first time using the realistic LQCD results of Ref.~\cite{Detmold:2015aaa}.

\begin{widetext}

\begin{figure}[tbh]
\includegraphics[height=5.cm]{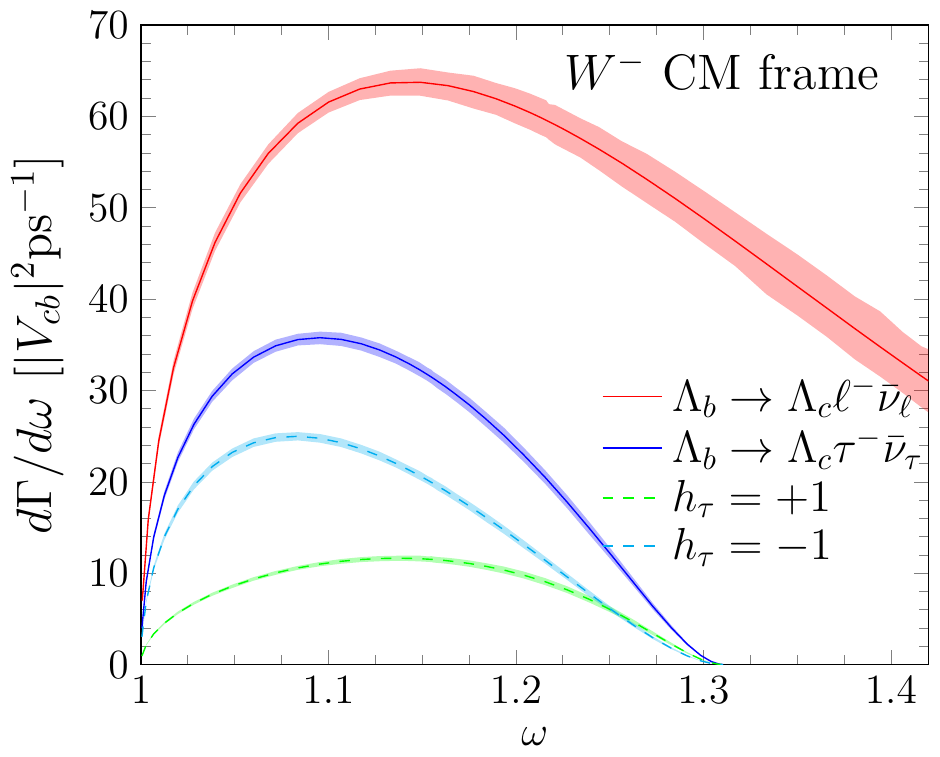}\hspace{1cm}
\includegraphics[height=5.cm]{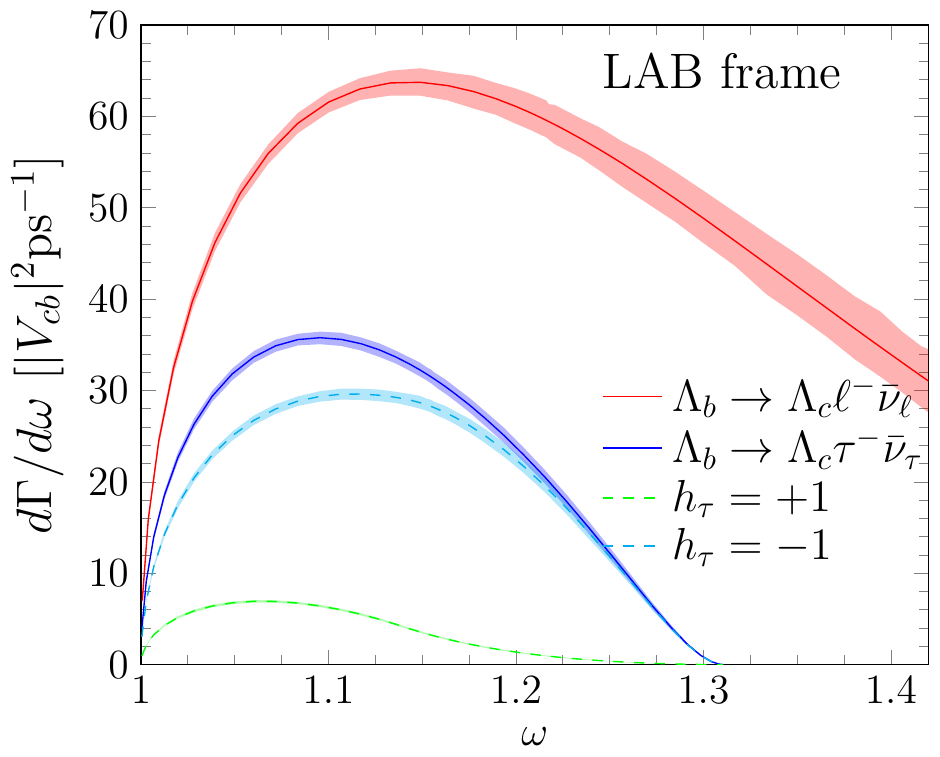}
\caption{  The $d\Gamma/d\omega$ distribution for the $\Lambda_b^0\to
  \Lambda_c^+$ semileptonic decay predicted using the LQCD form factors of Ref.~\cite{Detmold:2015aaa}. 
  The uncertainty bands account for 68\% CL intervals, and  here $\ell$ stands for a massless charged
  lepton. For the $\tau$ case we show the individual contributions from  tau leptons with positive ($h_\tau=1$) and 
  negative ($h_\tau=-1$) helicities,  measured in the $W^-$ CM  (left) and in the LAB (right) frames. }  
  \label{fig:dGdomega}
\end{figure}

\end{widetext}

For completeness,  in Fig.~\ref{fig:dGdomega} we show the $d\Gamma/d\omega$
differential decay width and its corresponding uncertainty band inherited from the statistical correlated fluctuations of the LQCD form-factors.  
For the $\tau$ case, we show  explicitly the SM predictions for the 
contributions from   tau leptons with positive and negative helicities, both in the CM and LAB frames. \\

\subsection{$c_2$ and $a_2$  sensitivity to NP} 
\label{sec:c2a2}

In this section we shall investigate the effect of NP on the $c_2$ and $a_2$ SFs  for the 
$\Lambda_b\to\Lambda_c$ semileptonic decay. The derived formulas are general and not specific to this transition. We shall consider  the  effective
Hamiltonian
\bea
&&\hspace{-.5cm}H=\frac{4G_F|V_{cb}|^2}{\sqrt2}[(1+C_{V_L}){\cal O}_{V_L}+
C_{V_R}{\cal O}_{V_R}+C_{S_L}{\cal O}_{S_L}\nonumber\\
&&\hspace{2.5cm}+C_{S_R}{\cal O}_{S_R}+C_{T}{\cal O}_{T}]+h.c.,
\label{eq:hnp}
\eea
taken from Ref.~\cite{Murgui:2019czp}. The Wilson coefficients, $C_i$,  
parametrize possible deviations from the SM, i.e. $C_i^{\rm SM}$=0, and could be in general, lepton
and flavour dependent, though in \cite{Murgui:2019czp} are assumed to be present only in the third generation
of leptons. Moreover, these LECs are taken to be real (CP-symmetry conserving limit). Complex Wilson coefficients can
explain the $b\to c\tau\nu_\tau$ anomalies as well as real ones, but they do not offer any clear advantages
regarding the fit quality, so they have not been considered in the effective low-energy Hamiltonian approach of Ref.~\cite{Murgui:2019czp} for simplicity. 
In Table 6 of that reference, the authors provide
four different fits (4, 5, 6 and 7) that include all the above terms.
Of these four fits, we shall only 
consider the last two. The reason being that for Fits 4 and 5 the SM coefficient 
is almost  canceled and its effect is replaced
by  NP contributions, what seems to be  an unlikely situation from a physical point of view.
In Fits 6 and 7 the $C_T$ Wilson coefficient is very small ($0.01^{+0.09}_{-0.07}$ and $-0.02^{+0.08}_{-0.07}$, respectively) and here for
simplicity we shall make it zero. With these approximations,
  the amplitude changes from the 
original current-current  $J_{H rr'\,\mu}(p,p')\,J_{L h}^{\mu}(k,k')$ term to
\bea 
\widetilde J_{Hrr'\,\mu}(p,p')
\,J_{L h}^{\ \ \mu}(k,k')+
J_{Hrr'}(p,p')\,J_{L h}(k,k')\,
\eea
where $r,r'$ and $h$ are the polarization indexes for the initial  and  final hadrons
  and final charged lepton,  respectively. Note that 
$J_{H rr'\,\mu}(p,p')$ is the CC current in the general definition of the SM hadron
 tensor in Eq.~\eqref{eq:defWmunu} (or in Eq.~(\ref{eq.FactoresForma}) for the $\Lambda_b \to \Lambda_c$ case) and 
$J_{L h}^{\ \ \mu}(k,k')=\bar{u}^{(h)}_{\ell}(\vec{k}^{\,\prime}\,)\gamma^\mu(1-\gamma_5)
v_{\nu_\ell}(\vec{k}\,)$ is the usual 
vector-axial leptonic current.
In turn, $\widetilde J_{H rr'\,\mu}(p,p')$ is 
identical to  $J_{H rr'\,\mu}(p,p')$ but with the vector and axial form factors 
corrected by the multiplicative factors $C_V=(1+C_{V_L}+C_{V_R})$ and  $C_A= (1+C_{V_L}-C_{V_R})$, respectively.  As for $J_{Hrr'}(p,p')$ and $J_{L t}(k,k')$ they
are given by
\bea
J_{Hrr'}(p,p') & = & \langle H_c; p',r'|\bar c(C_S-C_P\gamma_5)b| H_b; p, r\rangle \\
J_{Lh}(p,p') & = &\bar{u}^{(h)}_{\ell}(\vec{k}^{\,\prime}\,)(1-\gamma_5)
v_{\nu_\ell}(\vec{k}\,).
 \eea
with $C_{S,P}=(C_{S_L}\pm C_{S_R})$.  The
product of the lepton and hadron tensors is now changed to
\bea
L^{\mu\nu}(k,k';h)\,\widetilde W_{\mu\nu}(p,q)
&+&L(k,k';h)\,W(p,q)\nonumber\\&+&L_\mu(k,k';h)\,W^\mu(p,q)
\eea
with $\widetilde W_{\mu\nu}(p,q)$ constructed with the vector and axial form factors
modified by using the multiplicative factors $C_V$ and $C_A$ introduced  above.  In the above expression,  $h=\pm 1$ is the charged lepton helicity, with 
the new lepton terms given by\footnote{The polarized vector tensor $L_\mu(k,k';h)$ 
contains also an imaginary contribution, proportional to 
$h \epsilon_{\mu\alpha\beta\delta}k^{\prime \alpha}k^\beta s^\delta$ which vanishes 
exactly in the CM and vanishes upon contraction with the corresponding hadronic tensor in
the  LAB frame.}
%
\bea
L(k,k';h) &=& \left(k\cdot k' + h\, k\cdot s\right)/2 \\
L_\mu(k,k';h)&=&\frac{m_\ell}{2} k_\mu + \frac{h}{2 m_\ell} \left(k'_\mu k\cdot s-s_\mu k\cdot k'\right)
\eea
while using Lorentz covariance, the hadron new contributions can be expressed as 
\begin{eqnarray}
 W(p,q) &= & \overline\sum J_{Hrr'}(p,p')J_{Hrr'}^*(p,p') = W_{SP} \\
W^\mu(p,q) & = & \overline\sum \Big\{ \widetilde J^\mu_{Hrr'}(p,p')J_{Hrr'}^*(p,p')\nonumber \\
&+&  [\widetilde J^\mu_{Hrr'}(p,p')]^*J_{Hrr'}(p,p')\Big\} \nonumber \\ 
&=& W_{I1} \frac{p^\mu}{M} + W_{I2} \frac{q^\mu}{M} \\
\nonumber 
\end{eqnarray}
where we have introduced three new real scalar SFs, $W_{SP}, W_{I1}$ and $W_{I2}$,
that depend on $q^2$ alone. We readily obtain the NP corrections to the 
double differential decay width $d^2\Gamma/(d\omega d\cos\theta_\ell) $ in the $W^-$boson CM frame (Eq.~\eqref{eq:distr-ang}), 
\bea
a_0 &\to&  \tilde a_0 + \left[ \left(W_{SP}+ \frac{m_\ell}{M} W_{I2}\right) \frac{q^2}{M^2} +\frac{m_\ell M_\omega}{M^2} W_{I1}\right] \nonumber \\ 
a_1 & \to &  \tilde a_1 + \frac{m_\ell M'}{M^2} \sqrt{\omega^2-1} W_{I1} \nonumber \\
a_2 & \to & \tilde a_2
\eea
where the different $\tilde a_j$ are given by  Eq.(\ref{eq:as}) 
with the $W_{1,\cdots,5}$ SFs replaced by their $\widetilde W_{1,\cdots,5}$ counterparts that appear in  the Lorentz 
decomposition of $\widetilde W_{\mu\nu}(p,q)$.  The NP additional corrections to $(\tilde a_i-a_i)$ come only from $h=1$, this is to say, they vanish for
left helicity charged leptons.  Similarly,  the NP corrections to the 
double differential decay width $d^2\Gamma/(d\omega dE_\ell) $ in the LAB frame (Eq.~\eqref{eq:distr-lab}) read
\bea
c_0 &\to&  \tilde c_0 +  \frac{q^2-m_\ell^2}{M^2}\left(W_{SP}+ \frac{m_\ell}{M}W_{I2}\right)+ \frac{2m_\ell M_\omega }{M^2} W_{I1} \nonumber \\
c_1 & \to &  \tilde c_1 -\frac{2m_\ell }{M} W_{I1} \nonumber \\
c_2 & \to & \tilde c_2
\eea
where the $\tilde c_j$ are given by  Eq.(\ref{eq:distr-lab}) 
with the $W_{1,\cdots,5}$ SFs replaced by their corresponding $\widetilde W_{1,\cdots,5}$.

Neither $c_2$ nor $a_2$ are 
 modified by the left and right 
 scalar NP terms, being only sensitive to the left and right
 vector  corrections. Moreover, both of them are now proportional to $\widetilde W_2$, 
 and hence the relation of Eq.~\eqref{eq:a2c2} 
 still holds, in this limit where the tensor NP contributions have been neglected.
\subsubsection{Results for the $\Lambda_b \to \Lambda_c$ semileptonic transition}
In this case, we have
\be
J_{Hrr'}(p,p') =\bar{u}^{(r')}_{\Lambda_c}(\vec{p}^{\,\prime}\,)[ 
\,C_S F_S-
C_P F_P\gamma_5\,]
 u^{(r)}_{\Lambda_b}(\vec{p}\,) \label{eq:defSP}
\ee
where $F_S$ and $F_P$  are the scalar and pseudoscalar form factors that are directly
related (see  Eqs.~(2.12) and (2.13) of Ref.~\cite{Datta:2017aue}) to the  $f_0$ vector and $g_0$ axial  ones  given in \cite{Detmold:2015aaa}
. We thus have
\begin{eqnarray}
  W_{SP}&=& \frac12 {\rm Tr}\Bigg[\frac{\slashed{p}'+M_{\Lambda_c}}{2M_{\Lambda_c}}\left(C_S F_S-
C_P F_P\gamma_5\right) \nonumber\\
    &\times&  \frac{\slashed{p}+M_{\Lambda_b}}{2M_{\Lambda_b}}\left(C_S F_S+
C_P F_P\gamma_5\right)\Bigg] \label{eq:wsp}
\end{eqnarray}  
while the interference hadron tensor reads
\begin{widetext}
\begin{equation}
  W_{\mu}(p,q)= \frac12 {\rm Tr}\Bigg[\frac{\slashed{p}'+M_{\Lambda_c}}{2M_{\Lambda_c}}\widetilde{\cal O}_\mu\frac{\slashed{p}+M_{\Lambda_b}}{2M_{\Lambda_b}} \left(C_S F_S+
C_P F_P\gamma_5\right) + \frac{\slashed{p}'+M_{\Lambda_c}}{2M_{\Lambda_c}}\left(C_S F_S-
C_P F_P\gamma_5\right) 
\frac{\slashed{p}+M_{\Lambda_b}}{2M_{\Lambda_b}} \gamma^0\widetilde{\cal O}_\mu^\dagger\gamma^0
      \Bigg]\label{eq:WI12}
\end{equation}  
\end{widetext}
with $\widetilde{\cal O}^\mu=\Bigg(\sum_{i=1}^3 {\cal O}^\mu_i\widetilde{\cal F}_i\Bigg)
$ and $\widetilde{\cal F}_i=(C_V\,F_i-\gamma_5C_AG_i), \,i=1,2,3$.
Expressions for $W_{SP}, W_{I1}$ and $W_{I2}$ in terms of  $\widetilde F_{S,P}=C_{S,P}F_{S,P}$ (Eq.~\eqref{eq:defSP}), $\widetilde F_i=C_V F_i$ and  
$\widetilde G_i=C_A G_i$ are given in  Appendix~\ref{app:np}.
\begin{widetext}

\begin{figure}[tbh]
\includegraphics[height=5.cm]{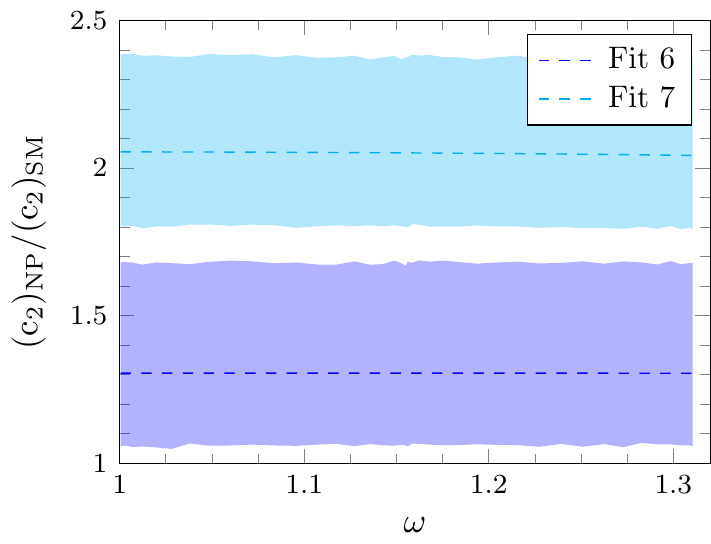}\hspace{1cm}
\includegraphics[height=5.cm]{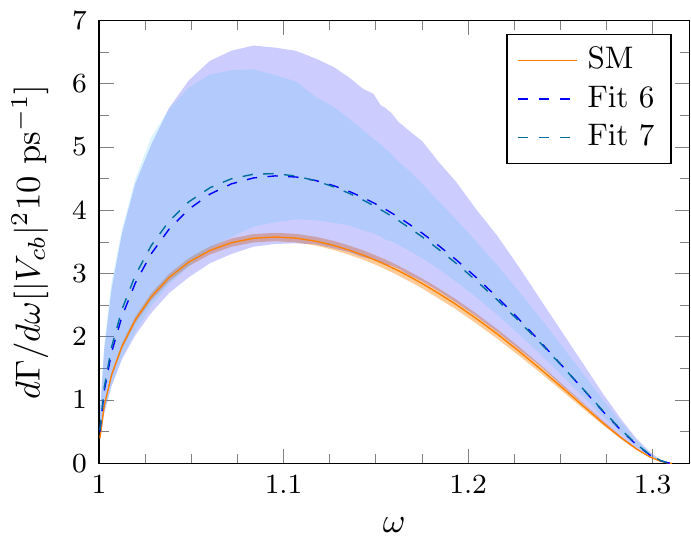}
\caption{Left: $(c_2)_{NP}/(c_2)_{SM}$ ratio for the $\Lambda_b\to
  \Lambda_c$ semileptonic transition obtained with the parameters of Fits 6 and 7  in Ref.~\cite{Murgui:2019czp}. Right: SM  and NP 
  predictions for the $d\Gamma/d\omega$ distribution for the $\tau$ decay mode.  As in other figures, the LQCD form factors of Ref.~\cite{Detmold:2015aaa} have been used and 
  the uncertainty bands account for 68\% CL intervals. }
  \label{fig:c2np}
\end{figure}

\end{widetext}

\begin{widetext}

\begin{figure}[tbh]
\makebox[0pt]{\includegraphics[height=4cm]{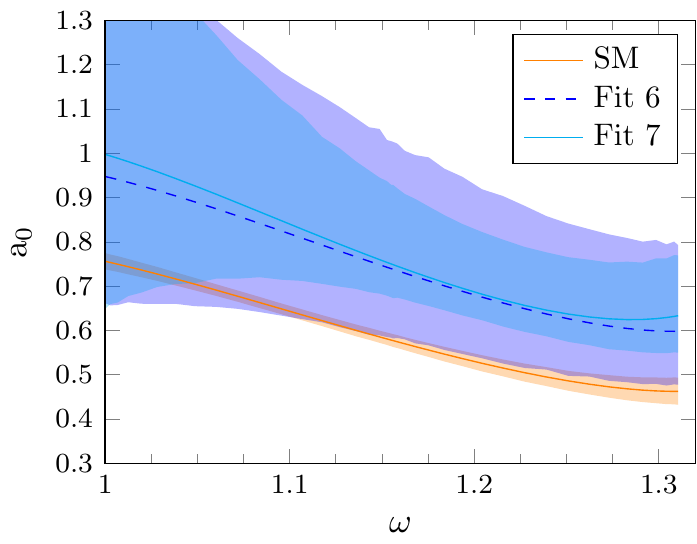} \includegraphics[height=4cm]{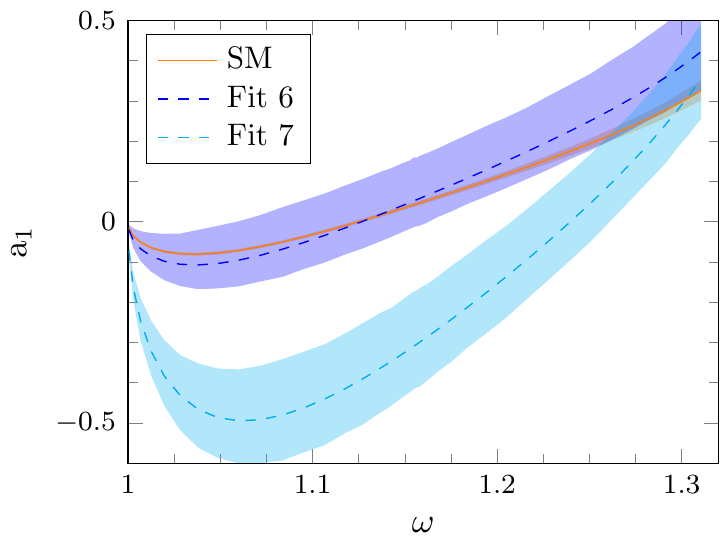}
\includegraphics[height=4cm]{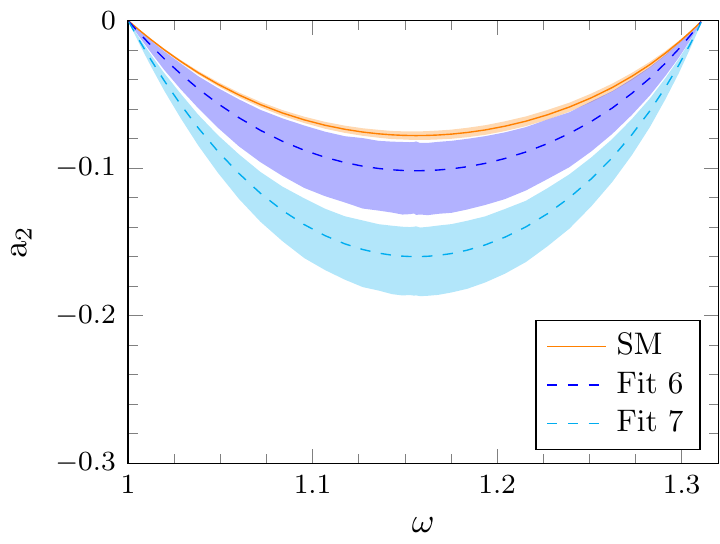}} \\
\makebox[0pt]{\includegraphics[height=4.1cm]{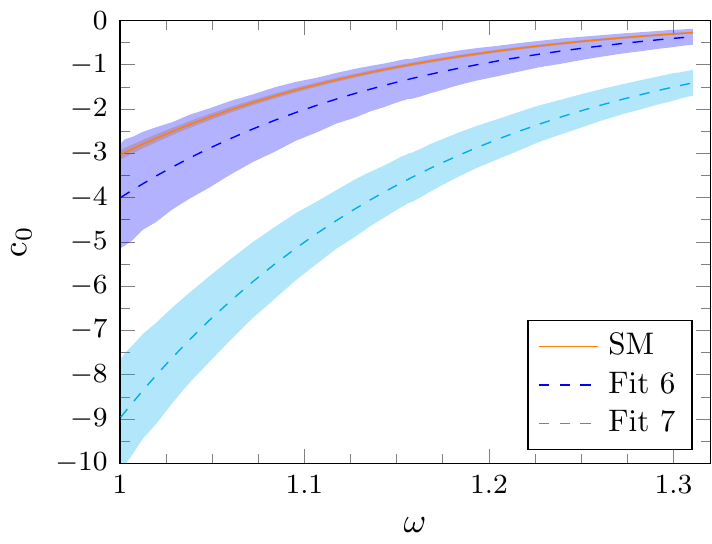} \includegraphics[height=4.1cm]{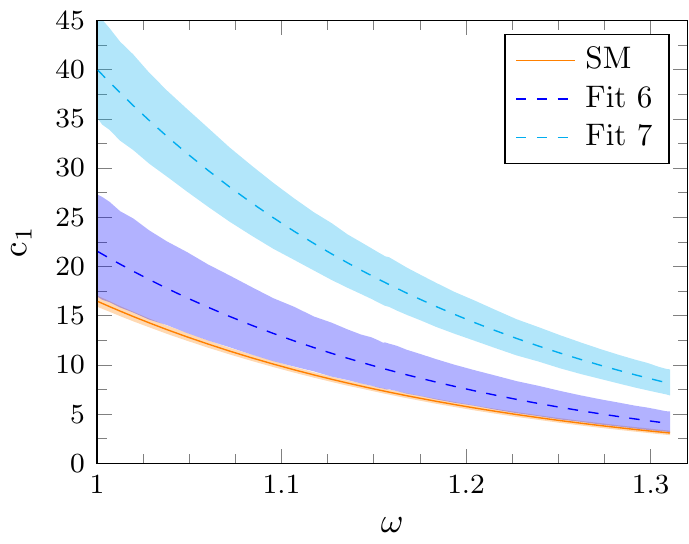} 
\includegraphics[height=4.1cm]{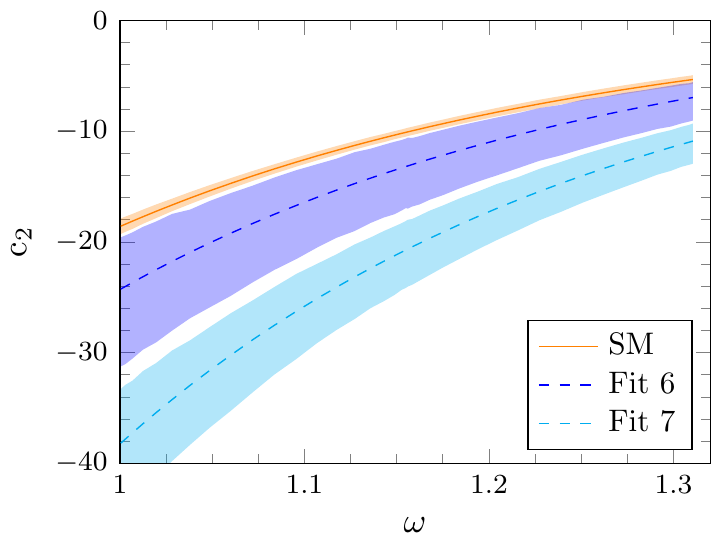}}
\caption{CM angular $a_0,a_1$ and $a_2$ (top) and LAB $\tau-$energy $c_0,c_1$ and $c_2$ (bottom) SFs
for the $\Lambda_b^0\to  \Lambda_c^+\tau^-\bar\nu_\tau$ decay obtained using the LQCD 
  form factors of Ref.~\cite{Detmold:2015aaa}. As in Fig.~\ref{fig:c2np},  we show the  SM predictions and the NP results obtained from  Fits 6 and 7  of Ref.~\cite{Murgui:2019czp}.}  
  \label{fig:a012yc012SmandNP}
\end{figure}

\end{widetext}

 As mentioned above, $c_2$ and $a_2$ are not affected by the left and right 
 scalar NP terms and these SFs are only modified by the left and right
 vector Wilson coefficients $C_{V_L,V_R}$. This turns out to be very relevant. Fits 
  6 and 7 in  Ref.~\cite{Murgui:2019czp} provide very different values for 
  $C_{V_L}$ and $C_{V_R}$ which implies different NP changes in  $c_2$ and
   $a_2$ SFs. However, the two fits produce very similar results for the
   ${\cal R}_{\Lambda_c}$ ratio (roughly 0.42, to be compared to the SM
   prediction of $0.33\pm 0.02$). In this situation the $c_2$ or
   $a_2$ SFs are observables that could differentiate  one fit from the other.
   In the left panel of Fig.~\ref{fig:c2np} we show the ratio $(c_2)_{NP}/(c_2)_{SM}=
   (a_2)_{NP}/(a_2)_{SM}$ as a   function of $\omega$ for the
   two fits under consideration. The $\omega$ dependence
   is hardly visible (for an explanation see the discussion below) but, as seen in the figure, the changes in magnitude of the NP corrections are significantly different in 
   the two fits, and are not accounted for by errors.  Hence, a measurement of $c_2$ for $\tau$ decay would not
   only be a direct measurement of the possible existence of NP, but it would also allow to
   distinguish from fits that otherwise give the same total and differential 
   $d\Gamma/d\omega$ decay widths (see right panel of Fig.~\ref{fig:c2np}).
   It would thus provide information on the  type of NP that is  needed to explain
   the data.

For completeness in Fig.~\ref{fig:a012yc012SmandNP}, we show the   
CM angular $a_0,a_1$ and $a_2$ (top) and LAB $\tau-$energy $c_0,c_1$ and $c_2$ (bottom) SFs as functions of $\omega$. In addition to $c_2$ and $a_2$, we find that $a_1$, 
and both $c_0$ and $c_1$ can also be used to distinguish between the two NP scenarios related to the minima 6 and 7 of
Ref.~\cite{Murgui:2019czp}. We recall here that the NP parameters were obtained in that reference from a general model-independent analysis of $b\to c\tau\bar\nu_\tau$ transitions, including measurements of ${\cal R}_{D}$, 
${\cal R}_{D^{*}}$, their $q^2$ differential distributions, the recently measured longitudinal  $D^*$ polarization $F_L^{D^*}$, and constraints from the 
$B_c \to \tau\bar\nu_\tau$ lifetime. We would like to stress that all $c_0,c_1$ and $c_2$ SFs, that determine the LAB 
$d^2\Gamma/(d\omega d E_\ell)$ distribution, are quite differently affected by the two NP settings analyzed here, even though both give rise to indistinguishable 
$d^2\Gamma/d\omega$ differential decay widths. 

Moreover, we also see that the ratio $(a_1)_{NP}/(a_1)_{SM}$  
would exhibit some sizable $\omega-$dependence, in particular in the case 
of Fit 7. This is in contrast to the case of the  
 $(c_2)_{NP}/(c_2)_{SM}$ ratio, depicted in the left panel of Fig.~\ref{fig:c2np}, 
 that turned out to be
 practically flat. This is because only linear  $C_{V_R}$ terms could induce a non-zero $\omega$ dependence for $(c_2)_{NP}/(c_2)_{SM}$, but
 however,
 to a high degree of aproximation (it would be exact in the heavy quark limit), 
 $\widetilde W_2$ is given by $M_{\Lambda_b}(\widetilde F_1^2(\omega) + 
 \widetilde G_1^2(\omega))/M_{\Lambda_c}$ with $F_1(\omega)\sim G_1(\omega)$ and,
 in this approximation linear effects on $C_{V_R}$ cancel exactly.

In the discussions  on Figs.~\ref{fig:c2np} and \ref{fig:a012yc012SmandNP} above, we have assumed uncorrelated Gaussian distributions for the Wilson coefficients  
$C_{V_L}$, $C_{V_R}$, $C_{S_L}$ and $C_{S_R}$ and have averaged the asymmetric errors quoted in Ref.~\cite{Murgui:2019czp}, since correlation matrices are not provided in that reference. This should be sufficient for the illustrative purposes of this subsection. Nevertheless, in what follows we will estimate the effects produced by the correlations between the Wilson coefficients in the $(c_2)_{NP}/(c_2)_{SM}$ ratio.

\begin{widetext}

\begin{figure}[tbh]
\makebox[0pt]{\includegraphics[height=5.1cm]{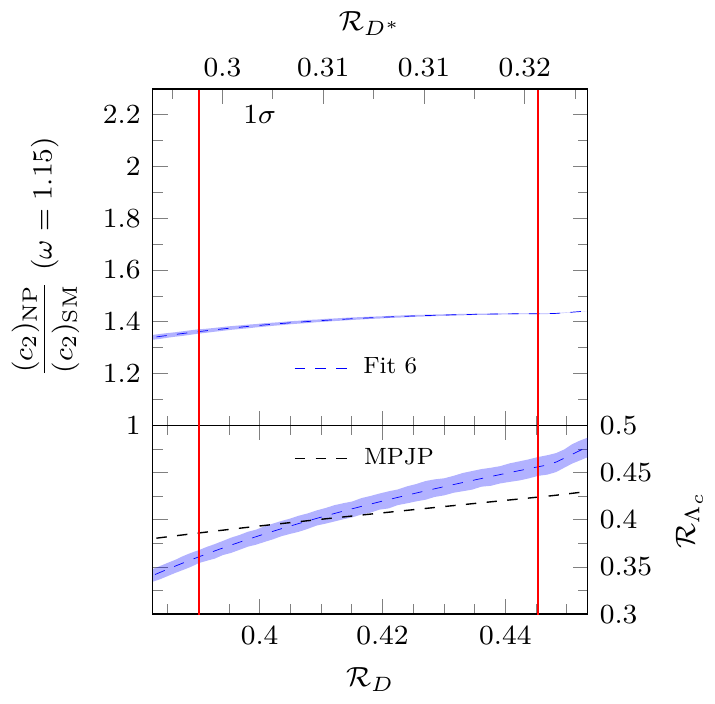} 
\includegraphics[height=5.1cm]{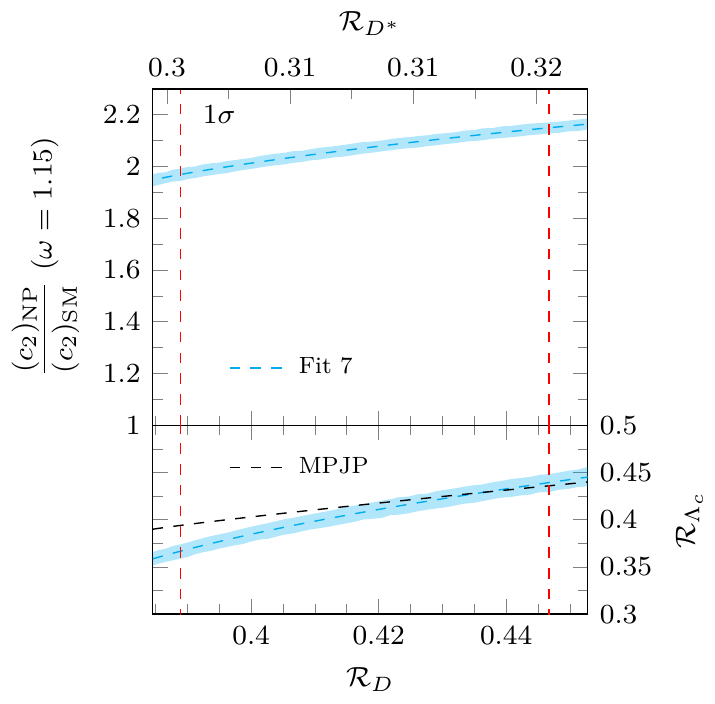}
\includegraphics[height=4.65cm]{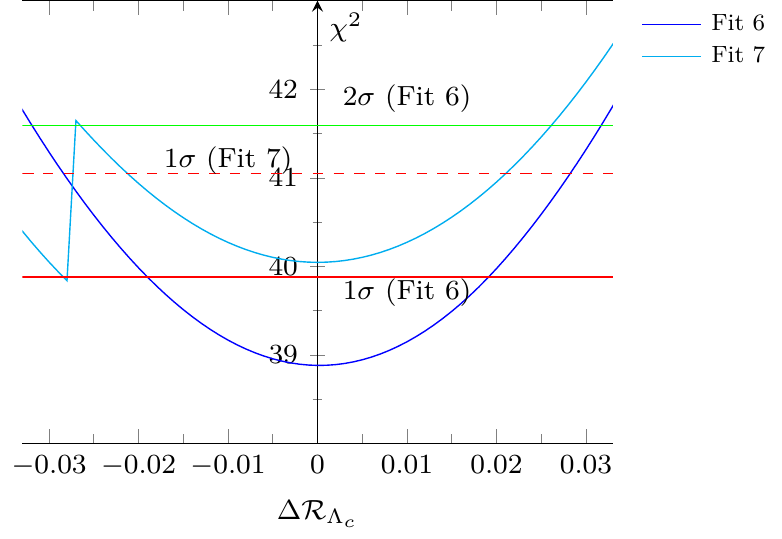}} \\
\caption{Left and middle panels: Effects of  Wilson-coefficients statistical  correlations on $(c_2)_{NP}/(c_2)_{SM}|_{\omega=1.15}$ for the NP  Fits 6 and 7  of Ref.~\cite{Murgui:2019czp}. Each set of Wilson coefficients is identified by its predictions for ${\cal R}_D$ and ${\cal R}_{D^*}$ ($X-$axes) and ${\cal R}_{\Lambda_c}$ (black dashed-curve, labelled as MPJP, in the bottom plots)~\cite{Murgui:2019czp,Ana}. SM predictions for the ${\cal R}_D^{\rm SM}=0.300 \pm 0.05$ and ${\cal R}_{D^*}^{\rm SM}=0.251 \pm 0.004$ ratios are below the ranges considered, while ${\cal R}^{\rm SM}_{\Lambda_c} = 0.33\pm 0.02$. We also show ${\cal R}_{\Lambda_c}$ computed after neglecting the NP tensor contribution (magenta and cyan dashed-lines for Fits 6 and 7, respectively). Shaded bands in our baryon results stand for  68\% CL uncertainties  inherited from LQCD 
inputs~\cite{Detmold:2015aaa}. Right panel: Fits 6 and 7 chi-square values~\cite{Murgui:2019czp,Ana}   
for each set of Wilson coefficients (sWC) used in the left and middle panels, and represented in this plot by $\Delta{\cal R}_{\Lambda_c}={\cal R}_{\Lambda_c}^{\rm sWC}-{\cal R}_{\Lambda_c}^{\min}$, with  ${\cal R}_{\Lambda_c}^{\min}= 0.405$ and 0.415 for Fits 6 and 7, respectively. See text for more details.  }  
  \label{fig:c2-corr}
\end{figure}

\end{widetext}

Note that in Ref.~\cite{Murgui:2019czp}, the uncertainties  
of a given parameter $y_i$ was determined as the shifts $\Delta y_i$ around the best-fit value $y^{\rm min}_i$ of that parameter, such that the minimization of $\chi^2|_{y_i=y^{\rm min}_i+ \Delta y_i}$ varying all remaining parameters in the vicinity of the minimum leads to an increase $\Delta\chi^2=1$. This procedure leads, in general, to asymmetric errors, 
and to non-Gaussian correlations that cannot be accounted for by a single matrix.  The effects of these correlations on $(c_2)_{NP}/(c_2)_{SM}$ are shown in Fig.~\ref{fig:c2-corr}. We have chosen this ratio because it hardly depends on $\omega$, and thus in Fig.~\ref{fig:c2-corr} we have fixed it to the intermediate value of 1.15. In the left (Fit 6) and middle (Fit 7) panels of this figure, we depict $(c_2)_{NP}/(c_2)_{SM}|_{\omega=1.15}$ and ${\cal R}_{\Lambda_c}$ for several sets of Wilson coefficients, which give rise to the ${\cal R}_D$ and ${\cal R}_{D^*}$ values given in the bottom and top  $X-$axes.

The ratios ${\cal R}_D$, ${\cal R}_{D^*}$ and ${\cal R}_{\Lambda_c}$ (black dashed-curves in the bottom plots),  and the $\chi^2$ shown in the right panel of Fig.~\ref{fig:c2-corr} have been computed as described in ~\cite{Murgui:2019czp}\footnote{The $\chi^2$ function is defined in Eq.~(3.1)  of that reference, and it is constructed using the meson inputs collected in Subsect. 2.3.} and have been obtained from the authors of that reference~\cite{Ana}. In Fig.~\ref{fig:c2-corr}, the Wilson-coefficients space is scanned starting from Fit 6 and 7 minima, through successive  small steps in the multi-parameter space leading to moderate merit-function enhancements and  ${\cal R}_{\Lambda_c}$ variations  (see the  right plot of Fig.~\ref{fig:c2-corr}). There exist one-to-one relations between each  set of Wilson coefficients (sWC) used in the left (Fit 6) and middle (Fit 7) panels of Fig.~\ref{fig:c2-corr} and the chi-square values or the variations $\Delta{\cal R}_{\Lambda_c}(={\cal R}_{\Lambda_c}^{\rm sWC}-{\cal R}_{\Lambda_c}^{\min})$ shown in the right plot of the figure. Note that at some point for $\Delta {\cal R}_{\Lambda_c} < -0.02$, the local Fit 7 collapses into Fit 6.

We see that Fit 6 and 7 chi-squares  grow from their local minimum values, and the $\Delta\chi^2=1$ and $\Delta\chi^2=2.71$ increments can be used~\cite{Murgui:2019czp} to determine the  68\% ($1\sigma$) and 90\% ($2\sigma$)   CL intervals of the NP predictions for ${\cal R}_{D}$, ${\cal R}_{D^*}$ and ${\cal R}_{\Lambda_c}$.  We also use here these $\chi^2$ variations to estimate the uncertainties on the results for  $(c_2)_{NP}/(c_2)_{SM}|_{\omega=1.15}$. In addition, the 68\% CL errors induced from the $\Lambda_b \to \Lambda_c$ LQCD form-factors~\cite{Detmold:2015aaa} are very small for this ratio, and are showed by the shaded  bands in both plots. Comparing the results depicted in Fig.~\ref{fig:c2np} with  the variation of $(c_2)_{NP}/(c_2)_{SM}|_{\omega=1.15}\left[=(a_2)_{NP}/(a_2)_{SM}|_{\omega=1.15}\right]$ between the $1\sigma$-vertical lines shown in Fig.~\ref{fig:c2-corr}, we observe that the inclusion of the Wilson-coefficients statistical correlations reduces the uncertainties  on  this ratio by factors of five and three for the predictions obtained from Fits 6 and 7, respectively. Thus, now we find that these two NP scenarios give rise to results for this latter observable separated by more than 5$\sigma$, $1.40\pm 0.04$ versus $2.06\pm 0.09$, despite predicting fully compatible ${\cal R}_D$, ${\cal R}_{D^*}$ and ${\cal R}_{\Lambda_c}$ integrated ratios. This discussion strongly reinforces our previous conclusions from  
Fig.~\ref{fig:c2np}.

A final remark concerns on the errors induced by neglecting the tensor NP contribution. In the bottom plots 
of the first two panels of Fig.~\ref{fig:c2-corr}, we compare for different sets of Wilson coefficients, the  predictions for ${\cal R}_{\Lambda_c}$ obtained from the full model of Ref.~\cite{Murgui:2019czp} (MPJP black dashed-curves) with those obtained in this work (magenta and cyan dashed-lines), where $C_T$ has been set to zero. For the latter predictions, we also display the errors (68\% CL bands) inherited from the $\Lambda_b \to \Lambda_c$ LQCD form-factors. We see that within the $1\sigma$ intervals, both for Fit 6 and 7, the effects of the NP tensor term on ${\cal R}_{\Lambda_c}$ are moderately small, and are partially accounted for the uncertainties of the LQCD inputs. This continues to be the case for all sets of Fit 7 Wilson coefficients considered in the $\chi^2-$plot of Fig.~\ref{fig:c2-corr}, while for Fit 6 and in regions above 1$\sigma$, $|C_T|$ appreciably grows and its effects become sizable.

\section{Summary}
We have introduced a general framework, valid for any $H\to H'
\ell\nu_\ell$ semileptonic decay, to study the lepton polarized CM
$d^2\Gamma/(d\omega d\cos\theta_\ell)$ and LAB
$d^2\Gamma/(d\omega d E_\ell)$ differential decay widths. To our knowledge,
 this is the first time  that the relevance of the $d^2\Gamma/(d\omega d E_\ell)$ differential decay width
has been put forward as a candidate for  LFU violation studies in $b\to c$
decays. Specifically, within the SM the $c_2(\omega)$ SF appearing in that distribution 
 is the same for all charged leptons. That makes it a perfect quantity
 for LFU violation studies. We have also found a correlation between the
 $a_2(\omega)$ SF related to the $(\cos\theta_\ell)^2$
 dependence in $d^2\Gamma/(d\omega d\cos\theta_\ell)$ and $c_2(\omega)$.
 This correlation 
 is shown in Eq.~(\ref{eq:a2c2}) and states that the ratio $a_2(\omega)/c_2(\omega)$, corrected by trivial
 kinematical and mass factors, gives a universal function valid for any
 $H\to H'$ semileptonic decay. Again, this is a clear prediction of the SM that
 can be checked against experiment.
  These two results could play a relevant role as further
  tests of the SM and LFU. 
  
We have also generalized the formalism to account for some NP terms, and shown that neither $c_2$ nor $a_2$ are 
 modified by left and right  scalar NP terms,  being however sensitive to  left and right
 vector  corrections. We also found that the relation of Eq.~\eqref{eq:a2c2} for the $a_2/c_2$ ratio 
 is not modified by these latter NP contributions.

 Finally, we have presented SM  and NP predictions for the $\Lambda_b \to \Lambda_c$ transition. We have shown  that a 
 measurement of $c_2$ (or $a_2$) for $\tau$ decay would not  only be a direct measurement of the possible existence of NP, but it would also allow to
 distinguish from NP fits to $b\to c\tau\bar\nu_\tau$ anomalies in the meson sector, that otherwise give the same total 
 and differential $d\Gamma/d\omega$  widths. The same applies to the other two SFs, $c_0$ and $c_1$, that appear in the LAB
$d^2\Gamma/(d\omega d E_\ell)$ differential width, and for the $a_1$ coefficient ($\cos\theta_\ell$ linear term) in the CM angular distribution.   

\section*{Acknowledgements}
We warmly thank F.J. Botella, C. Murgui, A. Pe\~nuelas  and A. Pich for useful
discussions. This research has been supported  by the Spanish Ministerio de
Econom\'ia y Competitividad (MINECO) and the European Regional
Development Fund (ERDF) under contracts FIS2017-84038-C2-1-P, FPA2016-77177-C2-2-P, 
SEV-2014-0398 and by the EU STRONG-2020 project under the program H2020-INFRAIA-2018-1, grant agreement no. 824093. 

\appendix
\begin{widetext}
\section{Hadron tensor SFs and form-factors for the $\Lambda_b^0\to  \Lambda_c^+\ell^-\bar\nu_\ell$ decay}
\label{app:ff}
The form factors used in Eq.~\eqref{eq.FactoresForma} are  related to those used in Ref.~\cite{Detmold:2015aaa} by
\begin{eqnarray}
F_1&=&f_{\perp},\nonumber\\
F_2&=&\frac{M_{\Lambda_b}\delta_{M_\Lambda}}{q^2}f_0+
\frac{M_{\Lambda_b}\Delta_{M_\Lambda}}{s_+}
\left[1-\delta\right]f_+-\delta_{s_{+}}f_{\perp},\nonumber\\
F_3&=&-\frac{M_{\Lambda_c}\delta_{M_\Lambda}}{q^2}f_0
+\frac{M_{\Lambda_c}\Delta_{M_\Lambda}}{s_+}
\left[1+\delta\right]f_+-\delta_{s_{+}}f_{\perp},\nonumber \\
G_1&=&g_{\perp},\nonumber\\
G_2&=&-\frac{M_{\Lambda_b}\Delta_{M_\Lambda}}{q^2}g_0-\frac{M_{\Lambda_b}\delta_{M_\Lambda}}{s_-}\left[1-\delta\right]g_+
-\delta_{s_{-}}g_{\perp},\nonumber\\
G_3&=&\frac{M_{\Lambda_c}\Delta_{M_\Lambda}}{q^2}g_0-\frac{M_{\Lambda_c}\delta_{M_\Lambda}}{s_-}\left[1+\delta\right]g_+
+\delta_{s_{-}}g_{\perp},\nonumber\\
\end{eqnarray}
with $\delta=(M_{\Lambda_b}^2-M_{\Lambda_c}^2)/q^2$, $s_{\pm}=(M_{\Lambda_b}\pm M_{\Lambda_c})^2-q^2$, 
$\delta_{M_\Lambda}=M_{\Lambda_b}-M_{\Lambda_c}$, $\Delta_{M_\Lambda}=M_{\Lambda_b}+M_{\Lambda_c}$ and $\delta_{s_{\pm}}=
2M_{\Lambda_b}M_{\Lambda_c}/s_{\pm}$.

On the other hand, from Eqs.~\eqref{eq:wmunu-baryons} and \eqref{eq:wmunu-dec}, we find

 \begin{eqnarray}
W_1&=& \frac{1}{2}\Big[(\omega-1) F_1^2+(\omega+1)G_1^2\Big],\nonumber\\
W_2&=&\frac{1}{2}\left\{ 2F_1F_2+(\omega+1)
F_2^2+2G_1G_2+(\omega-1)G_2^2+\frac{2M_{\Lambda_b}}{M_{\Lambda_c}}
\Big[(F_1+F_2)(F_1+F_3)+\omega(F_2F_3 +G_2G_3) \right.\nonumber\\
&+&\left.(G_1-G_2)(G_1+G_3)\Big]+\frac{M_{\Lambda_b}^2}{M_{\Lambda_c}^2}\Big[2F_1F_3
+(\omega+1)F_3^2+(\omega-1)G_3^2-2G_1G_3\Big]\right\},\nonumber\\
 W_3&=& \frac{2M_{\Lambda_b}}{M_{\Lambda_c}}F_1G_1,\nonumber\\
W_4&=& \frac{M_{\Lambda_b}^2}{2M_{\Lambda_c}^2}
\Big[ 2F_1F_3+(\omega+1)F_3^2+(\omega-1)G_3^2-2G_1G_3\Big],\nonumber\\
W_5&=&-\frac{M_{\Lambda_b}}{M_{\Lambda_c}}\Big[(F_1+F_2)(F_1+F_3)+\omega(F_2F_3+G_2G_3) +(G_1-G_2)(G_1+G_3)\Big]\nonumber\\
&-&\frac{M_{\Lambda_b}^2}{M_{\Lambda_c}^2}\Big[2F_1F_3 +(\omega+1)F_3^2
+(\omega-1)G_3^2-2G_1G_3\Big].\nonumber\\
\label{eq:ws}
\end{eqnarray}
%
%

\section{NP effects on the hadron SFs for the 
$\Lambda_b^0\to  \Lambda_c^+\ell^-\bar\nu_\ell$ decay}
\label{app:np}

From Eqs.~\eqref{eq:wsp} and \eqref{eq:WI12}, we find
\begin{eqnarray}
W_{SP}&=& \frac{1}{2}\Big[(\omega+1) \widetilde F_S^2+(\omega-1)\widetilde F_P^2\Big],\nonumber\\
W_{I1}&=& -W_{I2}+\widetilde F_S \left(\widetilde F_1+ (1+\omega)\widetilde F_2 \right) + \widetilde F_P\left(\widetilde G_1- (1-\omega)\widetilde G_2 \right),\nonumber\\
W_{I2}&=& \frac{M_{\Lambda_b}}{M_{\Lambda_c}}\left[\widetilde F_P\left(\widetilde G_1+ (1-\omega)\widetilde G_3 \right)- \widetilde F_S\left(\widetilde F_1+ (1+\omega)\widetilde F_3 \right)\right]
\label{eq:ws-np}
\end{eqnarray}
and $F_S= (M_{\Lambda_b}-M_{\Lambda_c})f_0/(m_b-m_c)$, $F_P= (M_{\Lambda_b}+M_{\Lambda_c})g_0/(m_b+m_c)$. In the numerical calculations, we use 
$m_b = 4.18\pm 0.04$ GeV and  $m_c = 1.27 \pm 0.03$ GeV as in Ref.~\cite{Datta:2017aue}.
\end{widetext}

\bibliography{B2Dbib}

\end{document}